\documentclass[11pt,a4paper]{article}
\usepackage{graphicx}
\usepackage{amsmath,amsfonts, epsfig} 
\usepackage{amssymb}
\usepackage{amscd}

\newcommand{\beqa}{\begin{eqnarray}}
\newcommand{\eeqa}{\end{eqnarray}}

\newcommand{\starco}[2]{\left[#1\stackrel{\star}{,}#2\right]}

\newcommand{\ri}{{\rm i}}

\renewcommand{\d}{\delta}

\renewcommand{\th}{\theta}
\renewcommand{\k}{\tilde{k}}

\newcommand{\F}{\widetilde{F}}
\newcommand{\D}{\widetilde{D}}

\newcommand{\wsq}{\widetilde{\square}}

\begin{document}

\title{Scalar and gauge translation-invariant noncommutative models}

\author{ A. TANASA\\
Department of Theoretical Physics\\
National Institute for Nuclear Physics and Engineering,\\
P.O.Box MG-6, RO-077125 Bucharest-Magurele, Romania,\\
E-mail: adrian.tanasa@ens-lyon.org}

\date{}
\maketitle
\begin{center}
(Received \today )
\end{center}

\begin{abstract}
We make here a short overview of the recent developments regarding translation-invariant models on the noncommutative Moyal space. A scalar model was first proposed and proved   renormalizable. Its one-loop renormalization group flow and parametric representation were calculated. Furthermore, a mechanism to take its commutative limit was recently given. Finally, a proposition for a renormalizable, translation-invariant gauge model was made.
\end{abstract}

{\centering\section{INTRODUCTION \label{intro} }}

Ever since noncommutative quantum field theory (at least on an Euclidean space) was proved (on the simplest noncommutative geometry, the Moyal space) to suffer from a new type of divergence, the UV/IR mixing \cite{mixing}, its renormalizability become the subject of intense studies.

The first  renormalizable scalar model on the Moyal space was the Grosse-Wulkenhaar (GW) model \cite{GW2}. Ever since several field theoretical methods were adapted within this noncommutative setting for the GW-like models (see for example \cite{dimreg}-\cite{param2} and references within). 

To obtain this perturbative renormalizability, to the propagator of the model was added a harmonic oscillator term. This idea was then extended to gauge theories on the Moyal plane \cite{gauge1}-\cite{gauge4}. 
Nevertheless, it does not seem easy to generalize this type of method to gauge theories: one is lead to theories with non-trivial vacua, in which renormalizability is unclear up to now.

The main drawback of the GW model is  the fact that it explicitly breaks translation-invariance. Furthermore, the fact that the harmonic term mentioned above does not seem to have a clear physical interpretation leads to difficulties in writing down some mechanism for its commutative limit (that is to recover the usual $\phi^4$ action when the noncommutative parameter $\theta$ of the Moyal space is switched to zero).

Answers to the above open questions were given by the proposal of another type of modification of the propagator. Leaving aside the GW harmonic term, in \cite{noi} the propagator was modified (in momentum space) with a $1/p^2$ term, term which is in fact explicitly given by the quantum corrections (at any loop order) of the noncommutative propagator - the effective propagator. This model was proved to be renormalizable at any order in perturbation theory \cite{noi}.

This new type of modification of the propagator was then adapted within an Euclidean  gauge theoretical context on the Moyal space \cite{gauge-noi}. This gauge action has a trivial vacuum and remains to see weather or not is renormalizable.

In this paper we make a short review of the recent field theoretical developments regarding these translation-invariant models. The paper is structured as follows. In the next section we recall the action of the model \cite{noi} and give some insights on its renormalization. The third section deals with its parametric representation  while the next one with its renormalization group flows. The fifth section presents the main idea of a commutative limit mechanism. The following section introduces the gauge models \cite{gauge-noi} based on this new type of modification of the propagator. Finally, the last section gives a few perspectives for future work.

{\centering\section{RENORMALIZABLE SCALAR MODELS}}

The Euclidean model defined in \cite{noi} on the $4$-dimensional Moyal space has action
\beqa
\label{revolutie}
S_\theta [\phi]=\int d^4 p (\frac 12 p_{\mu} \phi  p^\mu \phi  +\frac
12 m^2  \phi  \phi   
+ \frac 12 a  \frac{1}{\theta^2 p^2} \phi  \phi  
+ \frac{\lambda }{4!} V_\theta ),
\eeqa
where  $a $ is  some dimensionless  parameter and $V_\theta$ is the corresponding $\phi^4$ potential in momentum space.
The modified propagator is
\beqa
\label{propa-rev}
C(p,m,\theta) = \frac{1}{p^2+\mu^2+\frac{a}{\theta^2 p^2}} \, .
\eeqa
We further choose $a$ positive 
so that this propagator is positively defined.

The model \eqref{revolutie} mixes the UV and IR scales; it is this property which is responsible for its renormalizability.
The method used in \cite{noi} to prove this is the multi-scale analysis within the BPHZ renormalization scheme.
Let us recall here the following table summarizing the renormalization of the model  compared  to the one of the GW model and the ``one'' of the ``naive'' scalar model ({\it i. e.} the non-renormalizable model without the GW model or the $1/p^2$ term):
\begin{center}
\begin{tabular}{|l|c|c|c|c|c|c|}\hline

& \multicolumn{2}{|c|}{{\it ``naive'' model}}
& \multicolumn{2}{|c|}{{\it GW model}}
& \multicolumn{2}{|c|}{{\it  model \eqref{revolutie}}}\\ 
 & 2P & 4P & 2P & 4P & 2P & 4P\\
\hline
planar reg & ren & ren & ren & ren & ren & ren\\
\hline
planar irreg & UV/IR & log UV/IR & conv & conv & finite ren & conv \\
\hline
non-planar & IR div & IR div & conv & conv & conv & conv \\
\hline
\end{tabular}
\end{center}
where ``ren'' means renormalizable, ``conv'' (resp. ``div'')  means convergent (resp. divergent), ``reg'' (resp. ``irreg'') means regular (resp. irregular). For a definition of a regular Feynman graph one can report himself for example to \cite{param-GMRT}. Furthermore, ``2P'' and resp. ``4P'' mean $2-$point and resp. $4-$point Feynman graphs. We deal with them here because these are the graphs indicated to be primitively divergent by the power counting theorem proved in \cite{noi}. A similar power counting result was proved by different methods in \cite{param-GMRT}.

A minimalist version of the model \eqref{revolutie} was proposed and proved  renormalizable in \cite{fab}. Furthermore, the static potential associated to the model \eqref{revolutie} was calculated in \cite{altii}.

{\centering\section{PARAMETRIC REPRESENTATION}}

In this section we give some insights on the parametric representation of the model \eqref{revolutie}, representation obtained in \cite{param-GMRT}. The propagator \eqref{propa-rev} decomposes for $a<\theta^2 m^4/4$ as
\beqa
\label{propa2}
C(p,m,\theta)&=&\frac{1}{p^2+m^2}-\frac{1}{p^2+m^2}\frac{a}{\theta^2 p^2 (p^2+m^2)+a}
\nonumber\\
&=&
\frac{1}{p^2+m^2}-\frac{1}{p^2+m^2}\frac{a}{\theta^2 (p^2 +m_1^2)(p^2+m^2_2)},
\eeqa
where $-m_1^2$ and $-m_2^2$ are the roots of the denominator of the second term in the first line of the RHS (considered as a second order equation in $p^2$, namely
$\frac{-\theta^2 m^2\pm \sqrt{\theta^4 m^4 - 4 \theta^2 a}}{2\theta^2}<0$). 
The propagator further writes as
\beqa
\label{param}
&&C(p,m,\theta)=
\int_0^\infty d\alpha e^{-\alpha (p^2+m^2)}
\\
&&- 
\frac{a}{\theta^2} \int_0^\infty \int_0^\infty d\alpha d\alpha^{(1)} d\alpha^{(2)} e^{-(\alpha+\alpha^{(1)}+\alpha^{(2)})p^2} e^{-\alpha m^2} e^{-\alpha^{(1)} m_1^2}e^{-\alpha^{(2)} m_2^2}.\nonumber
\eeqa
Just as for the commutative $\phi^4$ theory, this decomposition allows to implement the parametric representation \cite{param-GMRT}.

{\centering\section{RENORMALIZATION GROUP FLOWS}}

In \cite{beta-GMRT} the one-loop renormalization group flows of the model \eqref{revolutie} are calculated. The computation relies on the decomposition \eqref{propa2}: the noncommutative correction does not contribute to the calculus of the $\gamma$, $\beta_m$ and $\beta_\lambda$ function at any order in perturbation theory. 
This correction is only responsible for the IR convergence of the integrals. 
Thus, these $\beta$ functions are computed as for the commutative $\phi^4$ model. The only difference is that only the planar regular graphs contribute to the renormalization of the mass, wave function and coupling constant. One has
\beqa
\beta_\lambda \propto \beta_\lambda^{{\rm commutative}}, \ \ \ 
\beta_m \propto \beta_m^{{\rm commutative}},\ \ \ 
\gamma \propto \gamma^{{\rm commutative}},
\eeqa
(see \cite{beta-GMRT} for the proofs). When regarding the new parameter $a$, its renormalization is finite, so one has
$$ \beta_a = 0.$$
Let us end this section by stating that in \cite{high} some higher order Feynman diagrams have been computed. The calculations do not use the decomposition \eqref{propa2} but the explicit Bessel function form of the amplitudes, thus not requiring any further condition on the positive parameter $a$.

{\centering\section{COMMUTATIVE LIMIT}}

In this section we present the main idea of the mechanism presented in \cite{limita} for obtaining the commutative limit of the model \eqref{revolutie}. Note that the new term in the action has a divergent, ``naive'' limit when $\theta\to 0$.

The strength of the mechanism proposed in \cite{limita} comes from the fact that the new term is directly dictated by the quantum correction of the propagator (at any loop order, as proved in \cite{limita}). When letting $\theta\to 0$ in this type of Feynman amplitude, one obtains the usual wave function and mass renormalization of commutative $\phi^4$ theory. Hence, when $\theta$ is turned off, the $1/(\theta^2 p^2)$ must not be present. 
One splits the usual mass and wave function counterterms (in the commutative $\phi^4$ theory) into two parts. The first is again a mass and wave function counterterm corresponding to the planar irregular graphs (present when $\theta\to 0$) and the second part is the $a$ counterterm (present only when $\theta\ne 0$).  
Taking all this into account (as well as the tricky behavior of all the other counterterms when $\theta$ is switched on and off) leads to a noncommutative action with associated counterterms which has a proper commutative limit (see \cite{limita} for further details).

{\centering\section{GAUGE MODELS}}

In this section we introduce the new gauge model of \cite{gauge-noi} (with the type of modification \eqref{revolutie}). 
Such a $U(1)$ gauge-invariant Euclidean action writes
\begin{align}\label{gauge}
\Gamma&=S_{\text{inv}}
+S_{\text{gf}}\,,
\\
S_{\text{inv}}&=\int d^4x\left[\frac 14 F^{\mu\nu}\star F_{\mu\nu}
+\frac 14 F^{\mu\nu}\star\frac{1}{D^2\tilde D^2}\star F_{\mu\nu}\right]
\,,
\nonumber\\
S_{\text{gf}}&=
\int d^4x\left[B\star
\left(1+\frac{1}{\square\wsq}\right)\partial^\mu A_\mu-\frac{\alpha}{2}B\star B
-\bar c\star\left(1+\frac{1}{\square\wsq}\right)\partial^\mu D_{\mu} c
\right]
\, .\nonumber
\end{align}
Here $\alpha$ is a real parameter, $c$ (resp. $\bar c$) is the ghost (resp. anti-ghost), $B$ is the Lagrange multiplier and 
\begin{align}
\F&=\th^{\mu\nu}F_{\mu\nu} \, , \qquad \ \ {\rm with} \ \;
F_{\mu\nu}=\partial_\mu A_\nu-\partial_\nu A_\mu
-\ri g\starco{A_\mu}{A_\nu}
\, ,
\nonumber\\
\D ^2 &= \D^{\mu} \star \D_\mu \, ,
\qquad {\rm with} \ \;
\D_\mu=\th_{\mu\nu}D^\nu
\,,
\end{align}
Note that 
$\frac{1}{D^2\tilde D^2}\star F_{\mu\nu}$ is to be understood as a formal power series in the gauge field $A_\mu$ (see \cite{gauge-noi} for further details).
Furthermore $1/\square$ denotes the Green function associated to $\square=\partial^\mu\partial_\mu=\sum_{i=1}^4 \partial_i^2$ and $\widetilde \square=\tilde \partial^\mu \tilde \partial_\mu,$ where $\tilde \partial \mu = \Theta_{\mu \nu}\partial^\nu$, $\Theta$ being the noncommutativity matrix  of the Moyal space.
In momentum space, one obtains the propagators
\begin{align}
G^{A}_{\mu\nu}(k)&=\frac{1}{k^2+\frac{1}{\k^2}}\left(-\d_{\mu\nu}
+\frac{k_\mu k_\nu}{k^2}-\alpha\frac{k_\mu k_\nu}{k^2+\frac{1}{\k^2}}\right)\,,
\nonumber\\
G^{\bar c c}(k)&=\frac{1}{k^2+\frac{1}{\k^2}}
\,.
\end{align}
In the Landau gauge $\alpha=0$, the $A$ propagator becomes
\begin{align}
G^{A}_{\mu\nu}(k)&=\frac{1}{k^2+\frac{1}{\k^2}}\left(-\d_{\mu\nu}
+\frac{k_\mu k_\nu}{k^2}\right)
\,. 
\end{align}

{\centering\section{PERSPECTIVES}}

It is a very interesting open question whether or not this type of noncommutative gauge theory is  renormalizable. Furthermore, from the point of view of particle physics of crucial importance is to investigate these models on a noncommutative Minkowski space.  
Note that some propositions already exist in the literature for approaching the construction of field theoretical models on such a space.
Some new insights could then be obtained for example in the sector of  Higgs physics.

{\centering\section{REFERENCES}}


\begin{center}

\end{center}
\end{document}